\documentclass{article}    

\usepackage{graphicx}

\title{\textbf{Objective vs Observer Measurements}}  
\author{Richard Mould\footnote{Department of Physics and Astronomy, State University of New York, Stony Brook,
\mbox{New York} 11794-3800; http://nuclear.physics.sunysb.edu/ \~{}mould}}  
\date{}    
   
\begin{document}             

\maketitle              

\begin{abstract}
Post-inflationary boundary conditions are essential to the existence of our highly structured universe, and these can only
come about through quantum mechanical state reductions -- i.e., through measurements.  The choice is between: An
`objective' measurement that allows reduction to occur independent of conscious observers, and an `observer' based
measurement that ties reduction to the existence of a conscious observer.  It is shown in this paper that that choice
cannot be determined empirically; so how we finally understand state reduction will be decided by the way that reduction
is used in a wider (future) theoretical framework.

\end{abstract}

\section*{Introduction}

 	In four previous papers\cite{RM1, RM2, RM3, RM4}, five rules of engagement are given that describe how
conscious observers engage quantum mechanical systems.  These rules do not include the Born rule (i.e., the Born
interpretation) that is replaced in this treatment by a rule that introduces probability through \emph{probability
current} only.  A summary of these rules and a more general discussion about them appears in another paper\cite{RM5}.  

The \emph{1st} rule describes a stochastic trigger that selects an eigenvalue of some variable on the basis of probability
current.  The \emph{2nd} rule is a selection rule that provides for the creation of `ready' brain states.  These are brain
states that are not initially conscious, but will become conscious the moment they are stochastically chosen.  The
\emph{3rd} rule says that when a ready brain state is stochastically chosen it will become conscious, and that all the
other components will go to zero.  The \emph{4th} rule is another selection rule that forbids a transition from a ready
brain state to another ready brain state of the same observer.  There is also a \emph{5th} rule that has no bearing on the
problem discussed in this paper.  

These rules provide for the collapse of a wave function in the presence of an observer.  This collapse is specifically
brought about by rule (3).

\vspace{0.3 cm}
\textbf{Rule (3)}: \emph{If a component that is entangled with a ready brain state B is stochastically chosen, then B will
become conscious, and all other components will be immediately reduced to zero.}
\vspace{0.3 cm}

Such luminaries as Von Neumann, Wigner, London, and Bauer have championed the idea that stochastic choice in quantum
mechanics depends on the presence of a conscious or potentially conscious observer as expressed by \mbox{rule (3)}. 
However, there is wide opinion among physicists that measurement (i.e., a stochastic choice) should take place independent
of the presence of a conscious observer.  It is easy to write a rule that would provide for such an `objective'
reduction.  I call it rule (1a).

\vspace{0.3 cm}
\textbf{Rule (1a)}: \emph{If a component of a superposition is locally incoherent with other components in the
superposition, and if it is stochastically chosen, then all those other components will be immediately reduced to
zero.}
\vspace{0.3 cm}

If the \emph{environmental decoherence} affecting any subsystem is sufficient to make one of its components locally
incoherent with other components, then according to rule (1a), a stochastic choice of that component will reduce all
those other components to zero.  A `state reduction' of this kind will be called an \emph{objective measurement} because
an observer need not be present for it to occur.  

 This rule is not itself a theory.  None of the rules referred to above constitute a theory.  They are only \emph{ad hoc}
claims about nature that potentially reflect the requirements of a wider, but still unknown, theoretical framework.  Rule
(1a) for instance, might be validated by a spontaneous reduction theory like that of Ghirardi, Rimini, and Weber
\cite{DG}, or a gravitational theory like that of Penrose \cite{RP}.  However, when considering rule (1a), I do not
assume that one of these theories is correct; and in fact, I am not inclined to believe either one of them.  That does not
prevent me from assuming that rule (1a) is correct on the basis of some as yet unknown theory, nor does it prevent me from
assuming that the other rules of engagement are correct in the same way.  In this paper, as in the previous papers, I work
only at the level of the rules themselves, without worrying about how they might be theoretically explained or justified. 
My goal is to get the right results (judged empirically) without regard to a rationale that might explain them.  I do this
by `grounding' the solutions of Schr\"{o}dinger's equation as completely as possible in observation.

\section*{Rules (1-4) are still Necessary}

Even if rule (1a) is adopted, we will still need rules (1-4).  Rule (1) will be required in any case in order to provide
for the existence of a stochastic trigger; and when there is an observer in the picture, rule (2) will be required to
insure the existence of ready brain states that perform the role described for them in refs.\ 1-5.  Rule (3) should be
modified in this case, for although it is necessary to insure that ready brain states become conscious when stochastically
chosen, it would be redundant for rule (3) to also require a collapse of the wave function.  So the adoption of rule (1a)
should be coupled with a modification of rule (3) called rule (3mod), where the latter provides only for the conversion of
a ready brain state to a conscious brain state upon stochastic choice.      Rule (4) will also be necessary in an
objective measurement scenario in order to prevent the appearance of the anomalous results described in refs.\ 1, 2, and
5.  

 An example of the effect of rule (1a) would be the reduction of the particle/detector system in eq.\ 1 in refs.\ 1 or 5. 
Prior to a stochastic hit at time $t_{sc}$ we would have

\begin{equation}
\Phi(t_{sc} > t \ge t_0) = \psi(t)D_0 + D_1(t)
\end{equation}
where the second component is zero at $t_0$ and increases in time.  The Schr\"{o}dinger process provides the required
probability current flow from the first component to the second component; and the two components will be incoherent
almost as soon as the second one is formed.  This insures an eventual stochastic hit on the second component according to
rule (1), because that rule associates positive current flow with the probability per unit time of a stochastic hit. 
Since there is no observer in this picture, rules (1) and (1a) will be entirely sufficient to affect a state reduction in
this \emph{objective measurement}, yielding

\begin{displaymath}
\Phi(t \ge t_{sc} > t_0) = D_1
\end{displaymath}
Any non-identity reduction is a \emph{boundary reduction} because it creates a new boundary for Schr\"{o}dinger's
equation.  See ref.\ 5 for a discussion of the importance of new boundaries in a post-inflationary universe.

Let an observer subsequently interact with $D_1$.  Following a stochastic hit at time $t_{sc(ob)}$ that occurs during the
ensuing physiological interaction, rules (1), (1a), (2), and    mod (3) will give rise to

\begin{equation}
\Phi(t \ge t_{sc(ob)} > t_{sc} > t_0) = D_1\underline{B}_1
\end{equation}
where the underline means that the brain state $\underline{B}_1$ is conscious.  

Now consider what will happen if rule (1a) is \emph{not in play} and rule (3) as originally stated is the sole source of
state reduction.  Then the detector in eq.\ 1 cannot by itself undergo state reduction, for an observer is not present. 
However, if a conscious observer is initially entangled with the detector, then the  interaction will
take the form given by eq.\ 2 in refs.\ 1 \mbox{and 5}.  

\begin{equation}
\Phi(t \ge  t_0) = \psi(t)D_0\underline{B}_0 + D_1(t)B_1
\end{equation}
where the second component is zero at $t_0$ and increases in time.  The underlined brain state $\underline{B}_0$ is
conscious, and the not-underlined brain state $B_1$ is a ready brain state.  Rule (2) is necessary at this point to insure
that the second component contains a ready brain state and not a conscious state.  Rule (2) is a selection rule that
prevents the brain's Hamiltonian from directly creating a discrete conscious state in this or in any interaction. 

 	Here again the current flow into the second component will cause a stochastic hit at some time $t_{sc}$. However, in
this case the resulting state reduction will be due to rule (3) as originally stated, for rule (1a) is not in play.  This
is also a boundary reduction that is  brought about by an \emph{observer measurement} yielding 

\begin{equation}
\Phi(t \ge  t_{sc} > t_0) =  D_1\underline{B}_1
\end{equation}
which is empirically indistinguishable from the objective measurement leading to eq.\ 2.  

Therefore, \emph{the observer cannot know if rule (1a) or rule (3) is correct}.  On purely empirical grounds we should
therefore abandon rule (1a) citing Occam's razor.  That's because we \emph{must} provide for a state reduction in the
presence of an observer, but it isn't empirically necessary to provide for a reduction otherwise.  However, this matter
will not be decided on empirical grounds.  In the end, the choice will be made on the basis of the theoretical arguments
that can be assembled on behalf of either kind of reduction.

\section*{A Terminal Observer}

	A \emph{terminal observer} is one who looks at the detector at a time $t_{ob}$ after the \mbox{time $t_f$} when the
particle/detector interaction has run its course.  This gives rise to a physiological interaction that alters eq.\ 1 as
shown in eq.\ 5 of ref.\ 1, giving
\begin{eqnarray}
\Phi(t \ge t_{ob} > t_f) &=& \psi(t)D_0X + D_1(t_f)X\\
&+& \psi'(t)D_0B_0 + D_1'(t_f)B_1\nonumber
\end{eqnarray}
where $X$ is the unknown brain state of the observer who does not initially interact with the detector.  Current flows from
the first row to the second row in eq.\ 5, where the components in the second row are equal to zero prior to $t_{ob}$. 
Since ready brain states are included in the only components that receive probability current, the stochastic choice that
is made in this case will be the same for an observer measurement as for an objective measurement.  So again, the observer
cannot tell if rule (1a) is or is not in effect.  The choice is empirically indeterminate.

\section*{Alternatives Empirically Unverifiable}

The above examples set the pattern.  In the appendix of this paper I go through all of the remaining cases studied in
refs.\ 1-2, and in all these cases the result is the same.  The observer cannot tell if rule (1a) is or is not the correct
reduction rule.  I conclude that \emph{it is impossible to empirically discriminate between an observer measurement
scenario and an objective measurement scenario.}  To quote a footnote in ref.\ 5,

\begin{quote}
``If an earlier boundary condition is uniquely correlated with a later observation, and if the observation is
stochastically chosen by the rules (1-3) proposed above, then it will be impossible for the observer to know if the
earlier boundary was also chosen by an earlier application of rule (1a).  This will be true even if the boundary is not
uniquely correlated with the observation.  Causal correlations that connect back to a range of possible earlier
`incoherent' boundaries will result in uncertainty in the boundary conditions that led to the observation.  So the precise
nature of the earlier boundary cannot be verified by the observer, and there will be no reason for him to believe that any
one of those possible boundaries was (or was not) the \emph{only} boundary by virtue of an earlier rule (1a) reduction. 
Therefore, to this extent, the effects of rule (1a) are empirically indistinguishable from the effects of rule (3).  An
exception may seem to be possible if causal connections go back to a range of earlier `coherent' boundaries; however, an
example like this is worked \mbox{out . . . with} results that also support indistinguishability."  
\end{quote}
This example is in the next section.

\section*{Nondemolition of  Coherent Boundaries}

	The discussion so far has assumed that the initial system is either a single state or an incoherent superposition or
mixture of states.  But the question is: What will happen if the system is initially a `coherent' superposition?  If the
result of a rule (3) observation depends on the properties of the superposition, then a rule (1a) reduction that occurs
prior to that observation might produce different results.  In that case, the observer would be able to tell if rule (1a)
is or is not in effect.  An example is the nondemolition measurement of a pair of 1/2 spin particles in a zero spin
state.  It should be possible to measure the total spin of this pair along any axis without measuring or interfering with
their total  spin along another axis, thus preserving the superposition.  However, a \mbox{rule (1a)} reduction might
occur when one of the two participating detectors makes contact with one of the particles, and that would destroy the
superposition.  It is shown below that that does not happen.  

Let $\Phi_1$  be the zero spin state $(J^2 = 0) = (2)^{-1/2}( \uparrow\downarrow - \downarrow\uparrow)$, plus a pair of
detectors.  The detectors are represented by a single symbol $D$ with two subscripts -- the first for the variable that
interacts with the first particle, and the second for the variable that interacts with the second particle.  Either
variable prior to the particle interaction is given by a subscript 0.

\begin{equation}
\Phi_I(\mbox{initially}) =  (2)^{-1/2}( \uparrow\downarrow - \downarrow\uparrow)D_{00}
\end{equation}

The two detectors are brought together at event \textbf{O} in fig.\ 1 so that their internal variables can be jointly
prepared in a (prescribed) way that leaves them entangled throughout the experiment.  The reason for this preparation will
be explained below.  The detectors separate after event \textbf{O}, allowing the first detector to interact with the first
particle at event \textbf{A} (in fig.\ 1).  This interaction occurs for a brief time between $t_1$ and $t_1 + \epsilon$,
giving  

\begin{eqnarray}
\Phi(t_1 + \epsilon > t > t_1) &=& c_1(t)\{(\uparrow\downarrow)D_{00} - (\downarrow\uparrow)D_{00}\}\\
&+& c_2(t)\{(\uparrow D_{10})\downarrow - (\downarrow D_{10})\uparrow\}\nonumber
\end{eqnarray}
where the second row in eq.\ 7 is zero at time $t_1$ and increases until time $t_1 + \epsilon$  when the first row goes
to zero.  The subscript 1 in $D_{10}$ means that the first detector has interacted with the first particle.   When this
interaction is complete, the state is

\begin{equation}
\Phi_{II}(t_2 > t > t_1 + \epsilon) =  (2)^{-1/2}\{(\uparrow D_{10})\downarrow - (\downarrow D_{10})\uparrow\}
\end{equation}

	I do not indicate a difference between the subscript 1 that appears in the first component of eq.\ 8 and the subscript 1
that appears in the second component.  It is true that these subscripts represent interactions with different spin
directions of the first particle; but the detector's internal variable is chosen in such as way that the interaction
causes it to become every bit as indefinite as the spin that it engages.  As a result, each $D_{10}$ in eq.\ 8 is
indefinite, so there is no possibility of distinguishing between the two states apart from the spin vector to which each
is correlated.  Therefore, the first detector cannot separately ``measure" spin up or spin down of the first particle in
this interaction; and since both detector states are indefinite, the entire function $\Phi_{II}$ is indefinite\cite{YA}.  

\begin{figure}[h]
\centering
\includegraphics[scale=0.8]{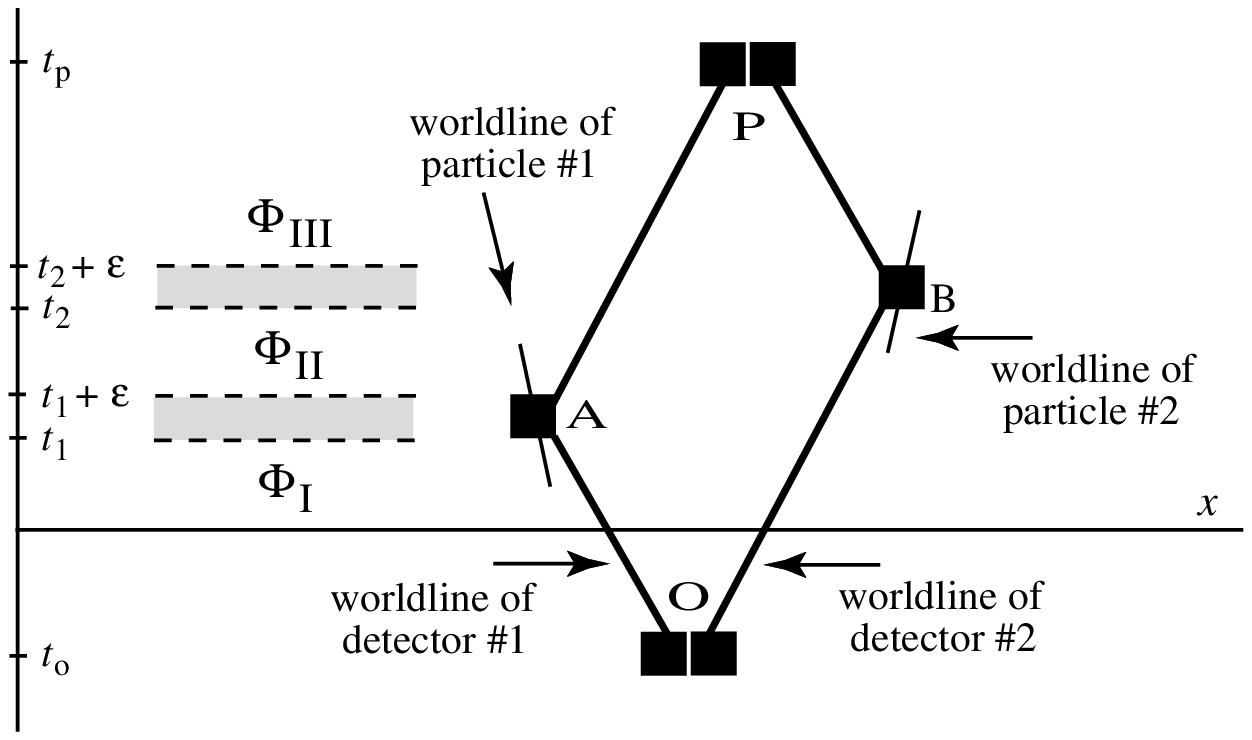}
\center{Figure 1}
\end{figure}

	Following event \textbf{A}, the second detector interacts with the second particle at   event \textbf{B} between times
$t_2$ and $t_2 + \epsilon$, giving 

\begin{eqnarray}
\Phi(t_2 + \epsilon > t > t_2) &=& c_3(t)\{(\uparrow D_{10})\downarrow - (\downarrow D_{10})\uparrow\}\\
&+& c_4(t)\{\uparrow D_{11}\downarrow - \downarrow D_{11}\uparrow\}\nonumber
\end{eqnarray}
where the second row is zero at time $t_2$ and increases until time $t_2 + \epsilon$  when the first row goes to zero.  The
final state of the system is therefore

\begin{displaymath}
\Phi_{III} = (2)^{-1/2}\{\uparrow D_{11}\downarrow - \downarrow D_{11}\uparrow\}
\end{displaymath}
which is the same as $(J^2 = 0)$ correlated with the final state of the detector pair.   

\begin{equation}
\Phi_{III} =  (2)^{-1/2}\{(\uparrow \downarrow - \downarrow \uparrow\}D_{11}
\end{equation}

Although the detector state in eq.\ 8 is indefinite, the interaction at \mbox{event \textbf{B}} restores the detectors to
a definite state $D_{11}$. This restoration is possible because of the initial joint preparation of the detector variables
at the time of \mbox{event \textbf{O}}.  That preparation correlates the internal variables in such a way that the
indefinite variable associated with the first detector (after event \textbf{A}) will add to the indefinite variable
associated with the second detector (after event \textbf{B}), to yield a definite variable associated with $D_{11}$.  See
ref.\ 8.  In this sequence, the detector pair has therefore made a definite nondemolition measurement of the particle pair,
finding them in the coherent superposition $(2)^{-1/2}(\uparrow \downarrow - \downarrow \uparrow )$ in which they began.  
Figure 1 shows the two detectors being physically reunited at event \textbf{P} so that their variables can be joined to
give a definite result.  

Now consider how state reduction might affect these results.

Let rule (3) be in play -- and \emph{not} rule (1a).  The observer will look at the detectors at the beginning (event
\textbf{O}) in order to verify the joint preparation of the internal variables of $D_{00}$.  He will again look at the
detectors at the end \mbox{(event \textbf{P})} in order to confirm the results of the measurement.  If the observer looks
at detector at event \textbf{A} (or \textbf{B}) alone, the result will be indefinite.  A state reduction at one of these
events would therefore destroy the superposition.  Definite results are observable only at the beginning and the end of
the experiment, and these preserve the superposition.  Therefore, an observation and state reduction will be allowed at
events \textbf{O} and \textbf{P}, but nowhere else.

On the other hand, imagine what will happen if rule (1a) is in play, \emph{rather than} the reduction feature of rule
(3).  A state reduction will certainly occur at events \textbf{O} and \textbf{P}, because a stochastic hit during the
detector/detector interaction will satisfy the requirements of rule (1a), as will a hit during the physiological
interaction with the observer.  The question is: Will rule (1a) also stochastically choose event \textbf{A} or
\textbf{B}, thereby reducing the other component to zero, and thereby demolishing the zero spin state?    

	Consider event \textbf{A}.  Rule (1a) reductions depend on a component of the super-position being locally incoherent
with other components in the superposition.  So the question is: Are the $D_{10}$ detector states appearing in each
component in eq.\ 8 locally incoherent; that is, are the environments associated with each of these detector states
orthogonal?  

The internal variables of the detectors appearing in the two components are correlated with one another, but they are
separately indefinite.  Their environments must not be allowed to interact with their respective internal variables, for
that would destroy this correlation.  As a result, the environment associated with one of the $D_{10}$ states in eq.\ 8 is
not measurably different than the environment associated with the other.  Consequently these environments are not
orthogonal, so the detector states in eq.\ 8 are not locally incoherent.  Therefore, the conditions for a rule (1a)
reduction are not met. The same might be said of the detectors emerging from event \textbf{B}.

It follows that even when rule (1a) has exclusive jurisdiction over reductions, a reduction will not occur at events
\textbf{A} or \textbf{B}.  However, there will be a reduction at events \textbf{O} and \textbf{P} because there will be a
rule (1a) stochastic hit during the detector/detector interaction in each case, if not during a subsequent physiological
interaction with the observer.  Here again, the observer cannot know if his final observation results from these
reductions due to rule (1a) or to rule (3).  The superposition is preserved in either case.    

In the previous section it was concluded that the distinction between an observer measurement scenario and an objective
measurement scenario is experimentally unverifiable.  This now appears to be true in the present example of initial
coherence, as well as in all the other cases considered in this paper and in the Appendix.

\section*{Nature of the Choice}

We have a choice between the objective and the observer measurement scenarios.  This is not just a matter of temperament,
for it is also a matter of judgment as to what the future holds.  There have been long suffering attempts to construct a
theory that supports the idea of objective state reduction.  None have been successful, and we have no reason to believe
that there is such a theory.  Nonetheless, many remain convinced that measurement must mean something independent of a
conscious observer.  

On the other hand, there is every reason to believe that there exists a theory of conscious brain states, for conscious
brains are clearly a functioning part of Nature.  There have been long suffering attempts in this direction too, but the
effort to understand consciousness and conscious brains is just beginning.  

It is my judgment that there is no general reduction theory independent of an observer\footnote{A footnote in ref.\ 5
makes an exception to this conclusion.  I don't think that superpositions of black holes are possible because of the
singularity at the center.  Therefore, it is entirely possible that the formation of a black hole automatically imposes a
boundary condition on the universe that is equivalent to an observerless state reduction.  }.  I believe that a proper
theory of consciousness will one day be found, and that it will provide the context in which the observer measurement
scenario will be vindicated.  A theory of that kind is surely necessary to validate the brainÕs selection rules (rules 2
and 4), and to address the part of rule (3) that changes a ready brain into a conscious brain. So either a future
\mbox{trans-Cartesian} theory will naturally include state reduction through rule (3), \emph{or} a separate theory will be
found whose task is to validate rule (1a).  I believe that the former is more likely.

\section*{Appendix}

In the following examples taken from previous papers, the empirical verifiability of rule (1a) of the objective
measurement scenario is compared with the empirical verifiability of  rule (3) of the observer measurement scenario. 

\vspace{.3 cm}
\noindent
\underline{An Intermediate Observer} (ref.\ 1, eqs.\ 7-9)

A stochastic hit on the third or fourth component in eq.\ 7 could be just as easily due to rule (1a) in the objective 
scenario, as to rule (3) in the observer scenario.  The same might be said of a follow-up hit on the third component that
leads to eq.\ 9.  There might also be a rule (1a) hit on the second component that results in a subsequent hit on the
fourth component; but since the observer cannot tell the difference between this and a direct hit on the fourth component,
rules (1a) and (3) are empirically indeterminate.

\vspace{.3 cm}
\noindent
\underline{An Outside Terminal Observer} (ref.\ 1, eq.\ 11)

	The fourth component cannot be chosen in this case because that would be an anomalous capture, inasmuch as rule (4) is in
effect in \emph{either} measurement scenario.  This means that only the third component in eq.\ 11 can be stochastically
chosen, and the observer cannot tell if the resulting reduction is due to rule (1a) or rule (3).  So again, the rules are
empirically indeterminate.

\vspace{.3 cm}
\noindent
\underline{An Intermediate Outside Observer} (ref.\ 1, eqs.\ 12-15)

There are four potential stochastic hits in the section covering eqs.\ 12-15, In every case, the resulting reduction might
be due to either rule (1a) or rule (3). Therefore, rules are empirically indeterminate.

\vspace{.3 cm}
\noindent
\underline{Drift Consciousness } (ref.\ 1, eq.\ 16)

The stochastic hit on any one of the ready brain components in eq.\ 16 might be due to either rule (1a) or rule (3). The
rules are therefore empirically indeterminate.

\vspace{.3 cm}
\noindent
\underline{Sequential Interactions with an Observer } (ref.\ 2, eqs.\ 2-5)

	There are three possible stochastic hits on the primed components in eq.\ 2, each of which might be due to either rule
(1a) or rule (3).  There might also be rule (1a) hits on either the second or third components that would result in
subsequent hits on the fifth or sixth components; and the observer could not tell the difference between these and direct
hits on components five or six.  The rules are therefore empirically indeterminate.

\vspace{.3 cm}
\noindent
\underline{Version I of Schr\"{o}dinger's Cat } (ref.\ 2, eq.\ 7)

	The second component might be stochastically chosen by either rule (1a) \mbox{or rule (3)}. The rules are therefore
empirically indeterminate.

\vspace{.3 cm}
\noindent
\underline{Version I with Outside Observer} (ref.\ 2, eq.\ 11)

	Stochastic hits on the second and third components might be due to either rule (1a) or to rule (3). All subsequent
hits on these components will be on components containing ready brain states, so they too might be due to either rule (1a)
or to rule (3).  It follows that the rules are empirically indeterminate.

\vspace{.3 cm}
\noindent
\underline{Version II of Schr\"{o}dinger's Cat} (ref.\ 2, eq.\ 12)

	A stochastic hit on the third component in eq.\ 12, might result from either rule (1a) or rule (3).    A rule (1a)
stochastic hit on the second component would result in an eventual hit on the third component, and the observer could not
tell the difference between this and a direct hit on the third component.  The rules are therefore empirically
indeterminate.

\vspace{.3 cm}
\noindent
\underline{Version II with Outside Observer} (ref.\ 2, eqs.\ 15-16)

	A hit on the fourth or fifth component of eq.\ 16 might be due to either rule (1a) or rule (3).  A primary plus a
follow-up hit on the third component of \mbox{eq.\ 16} will lead to eq.\ 15, and either one might be due to either rule
(1a) or rule (3).  There might also be a rule (1a) stochastic hit on the second component in \mbox{eq.\ 16} that would be
subsequently followed by a hit on the third or fifth component, but the observer could not tell the difference between
this and a direct hit on the third or fifth component.  The rules are therefore empirically indeterminate.  

\pagebreak
\vspace{.3 cm}
\noindent
\underline{Version II with a Natural Wake-Up} (ref.\ 2, eq.\ 19)

	Direct stochastic hits on either of the components containing $C$ or $C_N$ might come from either rule (1a) or rule (3). 
There might be a rule (1a) hit on the third component that would be subsequently followed by a hit on the fourth
component, but the observer could not tell the difference between this and a direct hit on the fourth component.  The
rules are therefore empirically indeterminate.


\begin{thebibliography}{99}

\bibitem{RM1}R.A. Mould, ``Consciousness: The rules of engagement", quant-ph/0206064
\bibitem{RM2}R.A. Mould, ``Schr\"{o}dinger's Cat: The rules of engagement",
 
quant-ph/0206065
\bibitem{RM3}R.A. Mould, ``Conscious Pulse I: The rules of engagement", 

quant-ph/0207005
   
\bibitem{RM4}R.A. Mould, ``Conscious Pulse II: The rules of engagement", 

quant-ph/0207165

\bibitem{RM5}R.A. Mould,``Quantum Brain States", \emph{Found. Phys.}, \textbf{33} No. 4, (2003),
quant-ph/0303064

\bibitem{DG}D, Giulini, et al, \emph{Decoherence and the appearance of a Classical World in Quantum Theory}, (Springer, 
New York, N. Y., 1996) p. 41-44

\bibitem{RP}R, Penrose, \emph{Shadows of the Mind}, (Oxford University Press, New York, \mbox{N. Y.}, 1994), p. 335 

\bibitem{YA}Y. Aharonov and D. Z. Albert, ``Can we make sense out of the measurement process in relativistic quantum
mechanics?", \emph{Phys. Rev. D}, \textbf{24} No. 2, \mbox{359-370}  (1981), sec. II


\end{thebibliography}
\end{document}